\documentclass[aps,pra,groupedaddress,reprint,amsmath,amssymb]{revtex4-1}

\usepackage{graphicx}% Include figure files
\usepackage{dcolumn}% Align table columns on decimal point
\usepackage{bm}% bold math
\usepackage[colorlinks,allcolors=blue]{hyperref}% add hypertext capabilities
\usepackage{physics}
\usepackage{comment}
\usepackage{natbib}
\usepackage{tikz}

\begin{document}

% Use the \preprint command to place your local institutional report
% number in the upper righthand corner of the title page in preprint mode.
% Multiple \preprint commands are allowed.
% Use the 'preprintnumbers' class option to override journal defaults
% to display numbers if necessary
%\preprint{}

%Title of paper
\title{Generating Robust Entanglement via Quantum Feedback}

% repeat the \author .. \affiliation  etc. as needed
% \email, \thanks, \homepage, \altaffiliation all apply to the current
% author. Explanatory text should go in the []'s, actual e-mail
% address or url should go in the {}'s for \email and \homepage.
% Please use the appropriate macro foreach each type of information

% \affiliation command applies to all authors since the last
% \affiliation command. The \affiliation command should follow the
% other information
% \affiliation can be followed by \email, \homepage, \thanks as well.

\author{Kensuke Gallock Yoshimura}
%\email[]{kensukegy@keio.jp}
%\author{Kenta Usui}
\author{Naoki Yamamoto}
%\email[]{Your e-mail address}
%\homepage[]{Your web page}
%\thanks{}
%\altaffiliation{}
\affiliation{
Department of Applied Physics and Physico-Informatics, Keio University,\\
Hiyoshi 3-14-1, Kohoku, Yokohama 223-8522, Japan}

%Collaboration name if desired (requires use of superscriptaddress
%option in \documentclass). \noaffiliation is required (may also be
%used with the \author command).
%\collaboration can be followed by \email, \homepage, \thanks as well.
%\collaboration{}
%\noaffiliation

\date{\today}

\begin{abstract}
Generating entangled states is one of the most important tasks in quantum information technology. 
However, in reality any entanglement generator must contain some characteristic uncertainty, and 
as a result the produced entangled state becomes an undesirable mixed state. 
This paper develops a coherent feedback control scheme that suppresses the characteristic 
uncertainty of a typical entanglement generator (non-degenerate optical parametric oscillator) for 
producing robust Gaussian entangled states. 
In particular, we examine a two-mode squeezed state and Gaussian four-mode cluster 
states to demonstrate the effectiveness of the proposed control method. 
\end{abstract}

% insert suggested PACS numbers in braces on next line
\pacs{}
% insert suggested keywords - APS authors don't need to do this
%\keywords{}

%\maketitle must follow title, authors, abstract, \pacs, and \keywords
\maketitle

%%%%%%%%%%%%%%%%%%%%%%%%%%%%%%%%%%%%%%%%%%%%%%%%
%%%%%%%%%%%%%%%%%%%%%%%%%%%%%%%%%%%%%%%%%%%%%%%%
%%%%%%%%%%%%%%%%%%%%%%%%%%%%%%%%%%%%%%%%%%%%%%%%

\section{Introduction}

Entangled states are an essential resource in quantum information technology, such as 
the quantum cryptography \cite{CryptographyAndEntanglement661} and the quantum 
teleportation \cite{Teleporting1895, Furusawa706, Teleportation.of.Braunstein.869}. 
In particular, continuous-variable (CV) systems are well-established platforms for 
demonstrating those quantum information processing \cite{QuantumInfo.Braunstein.513}; 
for instance, Gaussian CV cluster states \cite{Briegel2001, Continuous032318} is an 
important class of entangled states that can be applied to the one-way quantum computation 
\cite{Raussendorf2001}.

However, in practice there always exists a fragility issue in the process of generating 
entangled states, which as a result could largely degrade the performance of quantum 
information processing. 
To be specific, we here focus on the non-degenerate optical parametric oscillator (NDPO) 
\cite{NDPA1.1646, NDPA2.481, Clerk2010, Ou1992}, which can be used for generating 
various types of entangled states. 
The NDPO is an optical cavity containing a nonlinear crystal; two photons entering the 
nonlinear crystal will be amplified by a strong electromagnetic wave (pump), and at the 
same time these two photons become entangled; as a result the NDPO outputs an 
entangled light field. 
The fragility issue in this device is that the system parameters such as the pump gain and 
the cavity length easily change, which as a result induces fluctuation on the output entangled 
state. 
This means that the resulting entangled state must be a mixed state.

\begin{figure}[tp]
\centering
\includegraphics[width=\columnwidth]{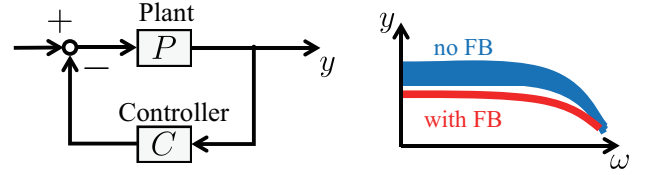}%
\caption{
Configuration of the general classical feedback control. 
When there is no feedback, the output $y$ fluctuates (see the right figure). 
By adding a feedback loop with controller $C$, the fluctuation of $y$ can be suppressed. 
}
\label{fig:EffectivenessOfFB}
\end{figure}

Therefore, it is important to devise a robust entanglement generator which is ideally free 
from the system's characteristic uncertainties. 
The key technique that is generally used for suppressing such a system fluctuation is 
feedback control. 
In the classical (non-quantum) case, the basic configuration of the feedback control is shown 
in Fig. \ref{fig:EffectivenessOfFB}. 
Let us consider a system $P$ (called the ``plant"), which outputs the signal $y$. 
For example, think $y$ as a voltage and $\omega$ as a frequency of the input. 
We want the voltage to be, say $y=5.0$ volts at $\omega=0$; 
however, due to the inevitable characteristic uncertainty contained in any electric device $P$, 
the output voltage must vary from the target value. 
The general solution is to feed a portion of the output $y$ back to the input by passing it 
through a robust system $C$ (called the ``controller"); 
then a suitably chosen controller may suppress the plant's fluctuation, and as a result the 
total system generates a less-fluctuating output. 
This effect can be clearly seen especially when the plant is given by an amplifier. 
Let us now interpret $P$ and $C$ as the gain at $\omega=0$ of the plant and the controller, 
respectively; 
then an input signal to the total system, $u$, is transformed to the output 
\[
      y=\frac{P}{1+PC}u,
\]
which converges to $y=u/C$ in the limit $P\rightarrow \infty$. 
This does not depend on $P$, and thus the output is robust against the plant's uncertainty 
involved in $P$.

Our idea is to apply the above idea to the problem of entanglement generation, where 
particularly the plant $P$ is given by a multi-mode Gaussian entanglement generator 
composed of a single NDPO and some beam splitters. 
In fact we show that, with the aid of feedback control, the controlled system becomes 
robust against the plant's fluctuation and obtains the ability to selectively produce 
several types of cluster states in a robust way. 
Note that this is a non-trivial extension of the work \cite{Yamamoto2016}, where a similar 
feedback scheme is applied to engineer a robust phase-insensitive quantum linear amplifier; 
here by the word ``non-trivial" we mean that how to configure the total feedback-controlled 
system composed of a multi-mode entangler and a controller is not clear, compared to 
the simple feedback amplification problem studied in \cite{Yamamoto2016}. 
Actually we show that, in the problem of generating four-mode cluster states, the effect of 
the feedback control differs depending on the structure of the feedback loop.

%%%%%%%%%%%%%%%%%%%%%%%%%%%%%%%%%%%%%%%%%%%%%%%%
%%%%%%%%%%%%%%%%%%%%%%%%%%%%%%%%%%%%%%%%%%%%%%%%
%%%%%%%%%%%%%%%%%%%%%%%%%%%%%%%%%%%%%%%%%%%%%%%%

\section{Preliminaries}
\label{sec:NDPOandFB}

%%%%%%%%%%%%%%%%%%%%%%%%%%%%%%%%%%%%%%%%%%%%%%%%
\subsection{The NDPO}
\label{sec:NDPO}

\begin{figure}[tp]
\centering
\includegraphics[width=6cm]{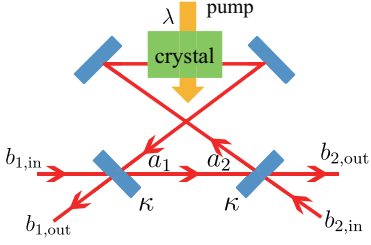}%
\caption{
Schematic of NDPO. 
The two input modes $b_{1,\mathrm{in}}$, $b_{2,\mathrm{in}}$ enter the system, and 
two outputs $b_{1,\mathrm{out}}$, $b_{2,\mathrm{out}}$ come out.
}
\label{fig:NDPA}
\end{figure}

Here we review the dynamics of the NDPO (see Fig.~\ref{fig:NDPA}). 
The NDPO has two internal modes (annihilation operators) ${a}_1$ and ${a}_2$, which are 
called the signal mode and the idler mode, respectively. 
The signal mode with frequency $\omega_1$ and the idler mode with frequency $\omega_2$ 
couple at the nonlinear crystal driven by the classical pump mode with frequency $2\omega_p$. 
The interaction Hamiltonian is given by 
\begin{equation}
{H}=\hbar \omega_1 {a}_1^\dag {a}_1 + \hbar \omega_2 {a}_2^\dag {a}_2
+i\hbar \lambda ( {a}_1^\dag {a}_2^\dag e^{-2i\omega_p t} - {a}_1 {a}_2 e^{2i\omega_p t}),
\notag
\end{equation}
where $\lambda \in \mathbb{R}$ is a coupling constant. 
The two modes $a_1$ and $a_2$ become entangled through this interaction; 
they leave the cavity and are transformed to an entangled traveling field between 
the field annihilation operators $b_{1,\mathrm{out}}$ and $b_{2,\mathrm{out}}$. 
Now let $b_{1,\mathrm{in}}$ and $b_{2,\mathrm{in}}$ be the input modes entering the cavity 
at the mirrors (see Fig.~\ref{fig:NDPA}). 
Then, the quantum Langevin equations \cite{Gardiner1985, gardinerQuantumNoise,wallsQuantumOptics} 
of the cavity modes in the rotating frame at $\omega_p$ are given by 
\begin{align}
\dfrac{d{a}_1}{dt}&=-\left( i\Delta_1+\frac{\kappa}{2} \right){a}_1+\lambda {a}_2^{\dagger} -\sqrt{\kappa}~{b}_{1,\mathrm{in}}, \label{eq:cavityEOM1} \\
\dfrac{d{a}_2^\dag}{dt}&=-\left( -i\Delta_2+\frac{\kappa}{2} \right){a}_2^{\dagger}+\lambda {a}_1 -\sqrt{\kappa}~{b}_{2,\mathrm{in}}^{\dagger},\label{eq:cavityEOM2}
\end{align} 
where $\kappa$ is a damping rate and $\Delta_j:=\omega_j - \omega_p$ are detunings of 
the cavity modes from the pump mode ($\hbar=1$ in this paper). 
Also, the input-output relations are
\begin{equation}
\label{eq:inputoutputrelation}
        {b}_{1,\mathrm{out}}=\sqrt{\kappa} {a}_1 + {b}_{1,\mathrm{in}},~
        {b}_{2,\mathrm{out}}^\dag=\sqrt{\kappa} {a}_2^\dag + {b}_{2,\mathrm{in}}^\dag.
\end{equation}
By Laplace transforming Eqs. \eqref{eq:cavityEOM1}, \eqref{eq:cavityEOM2} and 
\eqref{eq:inputoutputrelation}, followed by eliminating ${a}_1$ and ${a}_2$, one obtains
\[
      \left[\begin{array}{c}
         {b}_1(s)  \\
         {b}^{\dagger}_2(s)
      \end{array}\right]_{ \mathrm{out} }
      =G(s)
       \left[\begin{array}{c}
           {b}_1(s)  \\
           {b}^{\dagger}_2(s)
       \end{array}\right]_{ \mathrm{in} }, ~~
      G(s)=\left[\begin{array}{cc}
                 G_{11}(s) &G_{12}(s)  \\
                 G_{21}(s) &G_{22}(s) 
               \end{array}\right],
\]
where $s\in \mathbb{C}$ is the Laplace variable and $G_{ij}(s)$ are
\begin{align}
G_{11}(s)&=\frac{\left( s-\kappa/2+i\Delta_1 \right) \left( s+\kappa/2-i\Delta_2 \right)-\lambda^2}{D(s)}, \notag \\
G_{12}(s)&=G_{21}(s)=\frac{-\lambda \kappa}{D(s)}, \notag \\
G_{22}(s)&=\frac{\left( s+\kappa/2+i\Delta_1 \right) \left( s-\kappa/2-i\Delta_2 \right)-\lambda^2}{D(s)}.\notag
\end{align} 
with 
\begin{equation}
\label{eq:denomNDPO}
     D(s) := \left( s+\frac{\kappa}{2}+i\Delta_1 \right) \left( s+\frac{\kappa}{2}-i\Delta_2 \right)
                      -\lambda^2. 
\end{equation}
They satisfy 
$|G_{11}(i\omega)|=|G_{22}(i\omega)|$, 
$|G_{12}(i\omega)|=|G_{21}(i\omega)|$, and 
$G_{11}(i\omega)G_{22}(i\omega)-G_{12}(i\omega)G_{21}(i\omega)
=G_{22}(i\omega)/G_{11}^*(i\omega)$ for all $\omega$.

%%%%%%%%%%%%%%%%%%%%%%%%%%%%%%%%%%%%%%%%%%%%%%%%

\subsection{The quantum feedback amplification method}
\label{sec:CF for NDPO}

\begin{figure}[tp]
\centering
\includegraphics[width=\columnwidth]{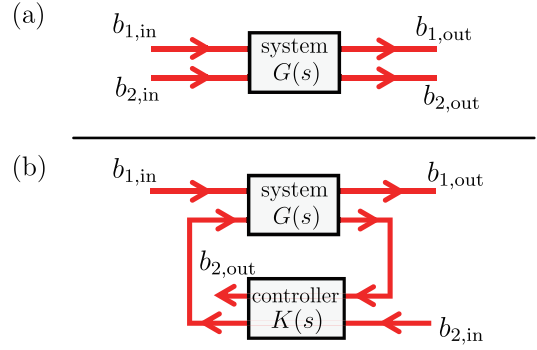}%
\caption{
(a) Diagram of the NDPO without control. 
(b) The coherent feedback configuration; 
the idler mode is used for feedback while the signal mode is not touched. 
}
\label{fig:WithWithoutFB}
\end{figure}

Here we review the coherent feedback method for engineering a robust quantum amplifier 
\cite{Yamamoto2016}. 
The plant $G$ is a general 2-inputs and 2-outputs linear phase-insensitive amplifier shown 
in Fig.~\ref{fig:WithWithoutFB}(a); the NDPO discussed above is a special class of this system. 
The feedback structure is shown in Fig.~\ref{fig:WithWithoutFB}(b); 
the idler output of $G$ is connected to the idler input of $G$, through a 2-inputs and 2-outputs 
linear passive quantum system $K$ such as an empty optical cavity, without involving any 
measurement process \cite{Wiseman4110, James1787, Gough, Hamerly173602,Kerckhoff021013,Yamamoto041029}. 
We express the controller's transfer function matrix $K(s)$ in the Laplace domain as
\begin{equation}
      K(s)=\left[\begin{array}{cc}
                  K_{11}(s) &K_{12}(s)  \\
                  K_{21}(s) &K_{22}(s) 
               \end{array}\right].\notag
\end{equation}
Note that, from the passivity property, $K(i\omega)$ is a unitary matrix satisfying 
$K(i\omega)K^\dag(i\omega)=I_2$ where $I_2=\mathrm{diag}\{1,1\}$. 
Then, the input-output relation of the total system depicted in Fig.~\ref{fig:WithWithoutFB}(b) 
is given by 
\begin{equation}
\left[
\begin{array}{c}
{b}_1(s)  \\
 {b}^{\dagger}_2(s)
\end{array}
\right]_{ \mathrm{out} }
=\left[
\begin{array}{cc}
G^{\mathrm{fb}}_{11}(s) &G^{\mathrm{fb}}_{12}(s)  \\
G^{\mathrm{fb}}_{21}(s) &G^{\mathrm{fb}}_{22}(s) 
\end{array}
\right]\left[
\begin{array}{c}
{b}_1(s)  \\
 {b}^{\dagger}_2(s)
\end{array}
\right]_{ \mathrm{in} }, \notag
\end{equation}
where
\begin{align}
G^{\mathrm{fb}}_{11}(s)
    &=\dfrac{G_{11}-K_{21}(G_{11}G_{22}-G_{12}G_{21})}{1-K_{21}G_{22}}, \notag \\%\label{eq:FBG11}
G^{\mathrm{fb}}_{12}(s)&=\dfrac{G_{12}K_{22}}{1-K_{21}G_{22}}, ~~
G^{\mathrm{fb}}_{21}(s)=\dfrac{G_{21}K_{11}}{1-K_{21}G_{22}}, \notag \\ %\label{eq:FBG21}
G^{\mathrm{fb}}_{22}(s)&=\dfrac{K_{12}+G_{22}\det K}{1-K_{21}G_{22}}.\notag %\label{eq:FBG22}
\end{align} 
From the equations below Eq.~\eqref{eq:denomNDPO}, one can obtain the following result:
\begin{align}
    |G^{\mathrm{fb}}_{11}(i\omega)|&=\left| \dfrac{G_{11}/|G_{11}|-K_{21}(G_{11}G_{22}-G_{12}G_{21})/|G_{11}|}{1/|G_{11}|-K_{21}G_{22}/|G_{11}|} \right| \notag \\ &\xrightarrow{|G_{11}|\rightarrow \infty} \dfrac{1}{|K_{21}(i\omega)|}.
\label{eq:FBfunction}
\end{align}
Therefore, if $G(i\omega)$ is a large-gain amplifier, the gain of the whole controlled system depends 
only on the passive (and thus robust) component $|K_{21}(i\omega)|$. 
That is, the controlled system functions as a robust amplifier with gain $1/|K_{21}(i\omega)|>1$.

%%%%%%%%%%%%%%%%%%%%%%%%%%%%%%%%%%%%%%%%%%%%%%%%
%%%%%%%%%%%%%%%%%%%%%%%%%%%%%%%%%%%%%%%%%%%%%%%%
%%%%%%%%%%%%%%%%%%%%%%%%%%%%%%%%%%%%%%%%%%%%%%%%

\section{Robust two-mode squeezed state}
\label{sec:RobustEPR}

In this section we show that the feedback amplification technique discussed in 
Sec.~\ref{sec:CF for NDPO} is effective for generating a robust two-mode squeezed (TMS) 
state. 
In our scenario this is a Gaussian entangled state between the signal and idler output fields 
of the NDPO \cite{Ou1992}. 
To quantify the entanglement of TMS state, we apply the {\it entanglement entropy}, which 
can be explicitly calculated in terms of the covariance matrix (CM) 
(see Appendix~\ref{sec:EntanglementMeasure}). 
The CM of the output state of the general (non-controlled) linear amplifier $G$ is given, 
in the frequency domain, by 
\begin{align}
    \gamma(i\omega)=\dfrac12
    \left[
\begin{array}{c|c}
(|G_{11}|^2 +|G_{12}|^2) I_2 & 2|G_{11}G_{12}|I_{1,1} \\ \hline
2|G_{11}G_{12}|I_{1,1} & (|G_{11}|^2 +|G_{12}|^2) I_2
\end{array}
\right],
\notag
\end{align}
where $I_{1,1}=\mathrm{diag}\{1,-1\}$. 
Also $|G_{11}|=|G_{22}|$ and $|G_{12}|=|G_{21}|$ are used. 
From this CM, one obtains the entanglement entropy $S(i\omega)$ of the TMS state as 
\begin{align*}
    S(i\omega) = |G_{11}(i\omega)|^2 \ln |G_{11}(i\omega)|^2 
                      - |G_{12}(i\omega)|^2 \ln |G_{12}(i\omega)|^2.
\end{align*}

Note that the output field state of the feedback-controlled amplifier is also a TMS state, and 
its covariance matrix $\gamma^{\mathrm{fb}}$ and the entanglement entropy 
$S^{\mathrm{fb}}$ can be obtained simply by replacing $G_{ij}$ by $G^{\mathrm{fb}}_{ij}$ 
in the above two equations. 
Then, as one can see, $\gamma^{\mathrm{fb}}$ and $S^{\mathrm{fb}}$ consist of 
$|G^{\mathrm{fb}}_{ij}|$, which is free from the characteristic uncertainty of the original 
amplifier $G$ if it has a large gain; 
as a consequence, the entanglement property of the output state of the feedback-controlled 
system also does not depend on those uncertainty. 
This is the central idea of robust entanglement generation via feedback amplification.

\begin{figure}[t]
\begin{center}
\includegraphics[width=8cm]{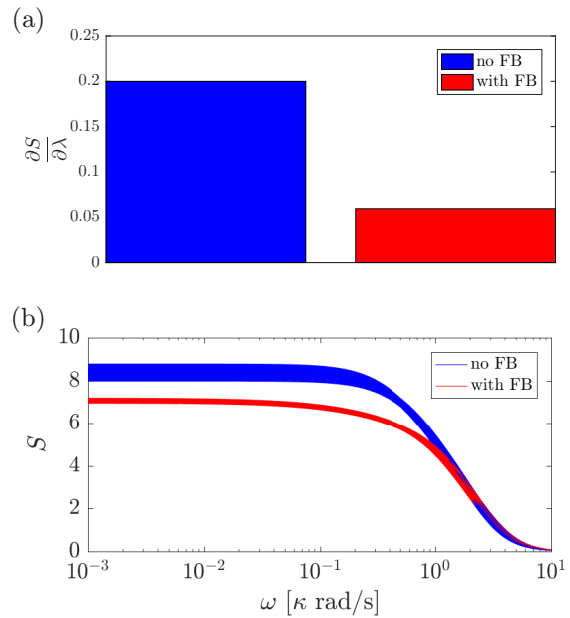}
\caption{
(a) The sensitivity of the entanglement entropy at $\omega=0$. 
(b) The entanglement entropy before (the blue lines) and after (the red lines) feedback as 
a function of $\omega$.}
\label{fig:EPRhist}
\end{center}
\end{figure}

We consider the NDPO discussed in Sec.~\ref{sec:NDPO}. 
The pole of this linear system is $s_{\pm}=-\kappa/2\pm\sqrt{\lambda^2-\Delta^2}$, and 
the gain at $s=0$ is 
\[
       |G_{11}(0)|^2=\frac{ \kappa^4 + 16(\Delta^2 - \lambda^2)^2 + 8\kappa^2 (\Delta^2 + \lambda^2) }
                                   {(\kappa^2-4\lambda^2+4\Delta^2)^2}, 
\]
where $\Delta_1=\Delta_2=\Delta$ is assumed. 
In the usual setting with cavity-locked NDPO ($\Delta=0$), the parameter is chosen as 
$\kappa\rightarrow 2\lambda+0$ to realize a high-gain amplification. 
However, this induces $s_+\rightarrow 0$, meaning that the amplifier becomes nearly unstable; 
that is,  there is a trade-off between the gain and stability of the system. 
To circumvent this issue, we take a special type of NDPO satisfying $\Delta=\lambda$ 
\cite{Yamamoto2016}; 
then, because $s_{\pm}=-\kappa/2$, such a trade-off does not appear. 
In this case, from $|G_{11}(0)|=\sqrt{1+16\lambda^2/\kappa^2}$, the gain of the NDPO 
increases monotonically with $\lambda$. 
Here the controller is set to a beamsplitter, which is independent of the frequency: 
\begin{equation}
\label{BS controller}
    K(i\omega)=
    K=\left[\begin{array}{cc}
            \tau &-\varrho  \\
            \varrho &\tau 
        \end{array}\right],~\tau^2+\varrho^2=1,
\end{equation}
where $\tau$ is the transmissivity and $\varrho$ is the reflectivity. 
In particular here we take $\varrho=0.04$, which satisfies the stability condition 
$|\varrho|<\kappa/2\lambda$ for the feedback-controlled system \cite{Yamamoto2016}. 
Also $\lambda=10\kappa$. 
The effect of the feedback can be clearly seen by examining the sensitivity of the entanglement 
entropy, $\partial S(i\omega)/\partial \lambda$; here, for simplicity, only the coupling strength 
$\lambda$ is assumed to change. 
Fig. \ref{fig:EPRhist}(a) shows the sensitivity of the non-controlled system 
$\partial S(0) / \partial \lambda$ and that of the controlled-one 
$\partial S^{\mathrm{fb}}(0) / \partial\lambda$. 
It is clear that the feedback control drastically lowers the sensitivity, meaning that 
the entangled state is robust against an unexpected change of $\lambda$.

\begin{figure}[t]
\begin{center}
\includegraphics[width=9cm]{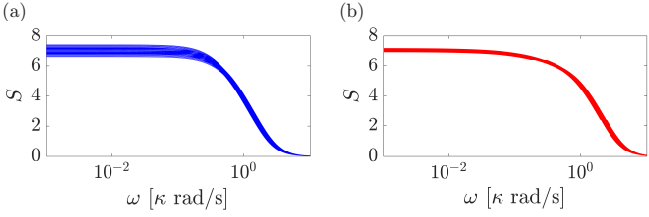}
\caption{
Entanglement entropy without (a) and with (b) feedback control. }
\label{fig:EPRcompare}
\end{center}
\end{figure}

Let us now see the robustness of the entanglement entropy in the frequency domain. 
The two parameters $\lambda$ and $\kappa$ can vary up to 10\% from their nominal values, i.e., 
$\lambda=(1+\delta_1)\lambda_0$ and $\kappa=(1 + \delta_2)\kappa_0$, where $\lambda_0$ 
and $\kappa_0$ are the nominal values satisfying $\lambda_0=10\kappa_0$. 
Figure~\ref{fig:EPRhist}(b) shows 90 samples of entanglement entropy, where the red and blue 
lines correspond to the case with and without feedback, respectively. 
Also $\delta_1$ and $\delta_2$ deterministically and linearly change from $-0.1$ to $0.1$. 
Clearly, in the low-frequency regime, the variation of $S^{\mathrm{fb}}(i\omega)$ due to the 
fluctuation of $(\lambda, \kappa)$ is smaller than that of  $S(i\omega)$, at the price of decreasing 
the degree of entanglement. 
It is noteworthy that this robustness property of the TMS state is provided intrinsically from 
the feedback control. 
This fact can be seen from Fig.~\ref{fig:EPRcompare}, comparing the non-controlled 
system satisfying $\lambda_0=5\kappa_0$ versus the feedback-controlled one satisfying 
$\lambda_0=10\kappa_0$, where in both cases the same 10$\%$ variations to these 
parameters as above are added; 
in fact the nominal values of $S(0)$ and $S^{\mathrm{fb}}(0)$ are nearly the same, but the 
variation of $S^{\mathrm{fb}}(0)$ is clearly smaller than that of $S(0)$.

%%%%%%%%%%%%%%%%%%%%%%%%%%%%%%%%%%%%%%%%%%%%%%%%
%%%%%%%%%%%%%%%%%%%%%%%%%%%%%%%%%%%%%%%%%%%%%%%%
%%%%%%%%%%%%%%%%%%%%%%%%%%%%%%%%%%%%%%%%%%%%%%%%

\section{Robust Gaussian cluster states}
\label{sec:RobustCluster}

In the previous section, we showed how the feedback control suppresses the fluctuation 
of the two-mode entangled state. 
Here we expand this result to a multi-mode case, in particular four-mode Gaussian cluster 
states with linear, T-shape, and square structures 
\cite{Continuous032318, Building032321, Ultracompact010302}. 
Although there are several ways to create cluster states \cite{Temporal062314}, we take 
the method using only a single multi-mode NDPO 
\cite{OneWay130501, Ultracompact010302, Zaidi2008, Analysis053826, Quadripartite063834}.

%%%%%%%%%%%%%%%%%%%%%%%%%%%%%%%%%%%%%%%%%%%%%%%%

\subsection{Linear cluster state}\label{sec:Linear}

\begin{figure}[t]
\begin{center}
    \includegraphics[clip,width=\columnwidth]{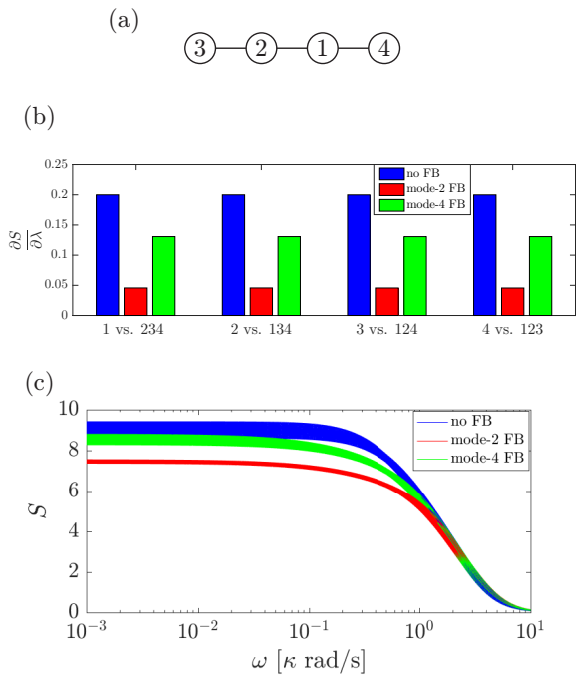}
\caption{
(a) Structure of the linear cluster state. 
(b) Sensitivities of the entanglement entropy. 
(c) Entanglement entropy between the mode-1 and the other modes, with (the red and green 
lines) and without (the blue lines) feedback.}
\label{fig:LinearCluster}
\end{center}
\end{figure}

We begin with a linear cluster state depicted in Fig.~\ref{fig:LinearCluster}(a), where 
the label $j~(=1,2,3,4)$ in the figure corresponds to $b_{j, \mathrm{out}}$. 
This is an output field state of a single multi-mode NDPO, as in the 
case of TMS state; see Appendix \ref{sec:ClusterAppendix}. 
The input-output relation of the NDPO in the Laplace domain is written as 
\begin{align}
\left[
\begin{array}{c}
b_1  \\
b_{2}^\dag  \\
b_{3}  \\
b_{4}^\dag 
\end{array}
\right]_{\mathrm{out}}
=G(s)
\left[
\begin{array}{c}
b_{1}  \\
b_{2}^\dag  \\
b_{3}  \\
b_4^\dag 
\end{array}
\right]_{\mathrm{in}}, \label{eq:In-OutAndPeak}
\end{align} 
where $G(s)$ is the $4\times 4$ transfer function matrix. 
Each matrix element contains the coupling constants $\lambda_l~(l=1,2,3)$, the damping 
rates $\kappa_j~(j=1,2,3,4)$, and the detunings $\Delta_j~(j=1,2,3,4)$. 
Here, for simplicity, we assume $\lambda:=\lambda_l$, $\kappa:=\kappa_j$, and 
$\Delta:=\Delta_j$ for all $l, j$.

\begin{figure}[t]
\centering
\includegraphics[width=6.5cm]{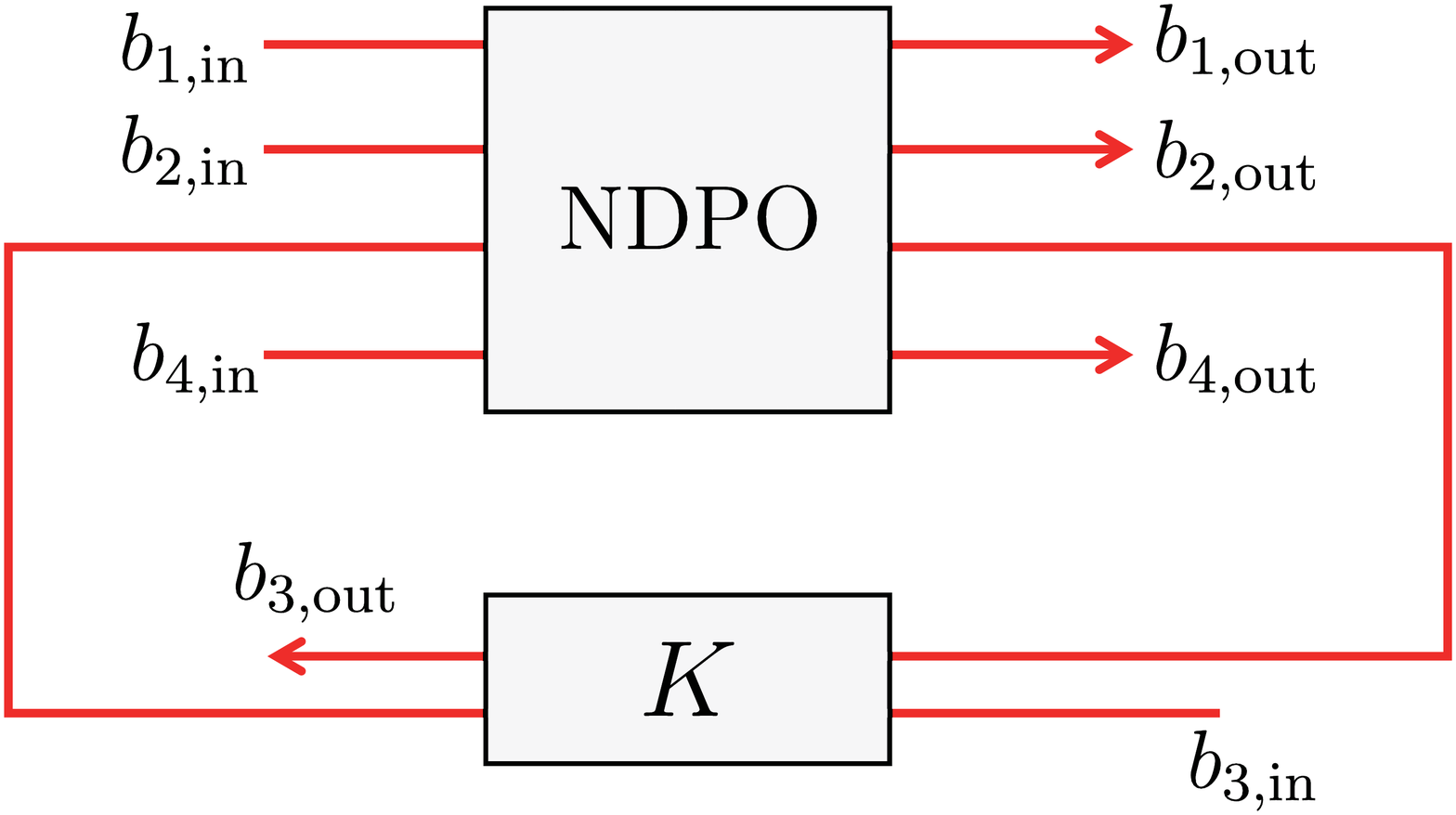}
\caption{
Feedback control where $b_{3}$ is used (the mode-3 FB).} 
\label{fig:FourModeFeedback}
\end{figure}

Let us apply the feedback control to the above 4 inputs and 4 outputs NDPO. 
Unlike Fig.~\ref{fig:WithWithoutFB}(b), the feedback configuration is non-trivial to design; 
we need to choose pairs of $(b_{j,\mathrm{out}}, b_{k,\mathrm{in}})$ and connect them via coherent 
feedback through a controller $K(s)$. 
In this paper we particularly consider {\it the mode-$j$ FB}, meaning that $b_{j,\mathrm{out}}$ 
and $b_{j,\mathrm{in}}$ are connected for a single index $j$ (Fig.~\ref{fig:FourModeFeedback} 
is the case of $j=3$). 
Then the transfer function matrix $G^{\mathrm{fb}}(s)$ for the feedback-controlled system is 
composed of the following elements:
\begin{align}
G_{jj}^{\mathrm{fb}}&=\dfrac{K_{12}+G_{jj}\det K}{1-G_{jj}K_{21}},\notag \\
G_{jk}^{\mathrm{fb}}&=\dfrac{G_{jk}K_{11}}{1-G_{jj}K_{21}},~ %(k \neq j),\notag \\
G_{lj}^{\mathrm{fb}}=\dfrac{G_{lj}K_{22}}{1-G_{jj}K_{21}},~(k, l\neq j),\notag \\
G_{lk}^{\mathrm{fb}}&=\dfrac{G_{lk}+K_{21}(G_{lj}G_{jk}-G_{jj}G_{lk})}{1-G_{jj}K_{21}},
~(k, l\neq j).\notag
\end{align}
Because $G_{jj}$ shows up in the denominators of $G^{\mathrm{fb}}$, the effect of feedback 
control appears when $|G_{jj}|\rightarrow \infty$. 
Now the poles of the NDPO is given by 
\begin{equation*}
    s = -\dfrac{\kappa}{2}\pm \sqrt{ \dfrac{3-\sqrt{5}}{2}\lambda^2 -\Delta^2 },~
          -\dfrac{\kappa}{2}\pm \sqrt{ \dfrac{3+\sqrt{5}}{2}\lambda^2 -\Delta^2 }.
\end{equation*}
As discussed in Sec.~\ref{sec:RobustEPR}, we take $\Delta=\sqrt{(3+\sqrt{5})/2}\lambda$, 
leading that all the poles strictly locate in the left-side of the complex plane for arbitrary large 
gain of the amplifier; that is, the gain-stability trade-off does not appear.

Let us now see the robustness property of the feedback-controlled system, by examining 
the entanglement entropy and its sensitivity. 
Again the beamsplitter controller \eqref{BS controller} with $\varrho=0.04$ is taken. 
Figure~\ref{fig:LinearCluster}(b) shows the sensitivities with $\lambda=10\kappa$ at $\omega=0$ 
for two cases of feedback way: the mode-2 FB (red bars) and the mode-4 FB (green bars). 
Clearly both of the feedback schemes reduce the sensitivity, but the degree of suppression 
differs depending on which mode is used for feedback. 
That is, the mode-2 FB makes the cluster state more robust than the mode-4 FB. 
This is because the mode-2 has two links while the mode-4 has only one; 
hence, the use of the former as feedback would be more effective to suppress the fluctuation 
added on all nodes. 
This difference can be observed in the frequency domain as well; 
Fig. \ref{fig:LinearCluster}(c) shows 90 sample paths of the entanglement entropy between 
the mode-1 and the other modes, where, as in the previous case, $(\lambda, \kappa)$ 
deterministically and linearly change up to 10$\%$ from their nominal values satisfying 
$\lambda_0=10\kappa_0$.

%%%%%%%%%%%%%%%%%%%%%%%%%%%%%%%%%%%%%%%%%%%%%%%%

\subsection{T-shape cluster state}

\begin{figure}[t]
\begin{center}
    \includegraphics[clip,width=\columnwidth]{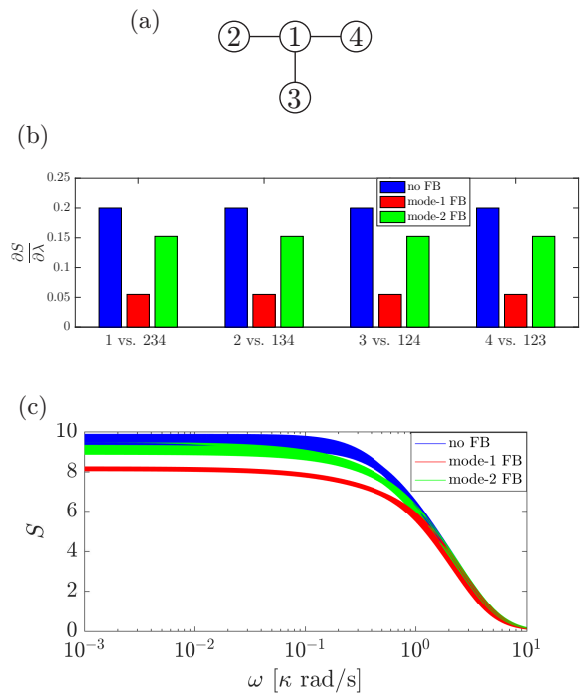}
\caption{
(a) Structure of the T-shape cluster state. 
(b) Sensitivities of the entanglement entropy. 
(c) Entanglement entropy between the mode-1 and the other modes, with (the red and green 
lines) and without (the blue lines) feedback.}
\label{fig:TshapeCluster}
\end{center}
\end{figure}

The next example is the T-shape cluster state whose structure is shown in 
Fig. \ref{fig:TshapeCluster}(a). 
The system Hamiltonian is given in Appendix \ref{sec:ClusterAppendix}, where for simplicity 
$\lambda=\lambda_l$, $\kappa=\kappa_j$, and $\Delta=\Delta_j$ are assumed. 
The poles of the non-feedback NDPO are $s=-\kappa/2 + i\Delta$ and 
$s=\kappa/2 \pm\sqrt{3\lambda^2-\Delta^2}$. 
Similar to the previous cases, we take $\Delta=\sqrt{3}\lambda$ to avoid the gain-stability 
trade-off. 
As for the controller, the beamsplitter with $\varrho=0.024$ is chosen. 
Under this condition, the sensitivity and the entanglement entropy are depicted in 
Figs.~\ref{fig:TshapeCluster}(b) and (c), where $\lambda_0=10\kappa_0$ and up to 10$\%$ 
fluctuation are added to $(\lambda, \kappa)$. 
Likewise the linear cluster case, the mode-1 FB scheme realizes the better suppression than 
the mode-2 FB, presumably because controlling the mode-1 can affect on all the other modes 
through the direct links while the mode-2 can do that only via an indirect way.

%%%%%%%%%%%%%%%%%%%%%%%%%%%%%%%%%%%%%%%%%%%%%%%%

\subsection{Square cluster state}

\begin{figure}[t]
\begin{center}
    \includegraphics[clip,width=\columnwidth]{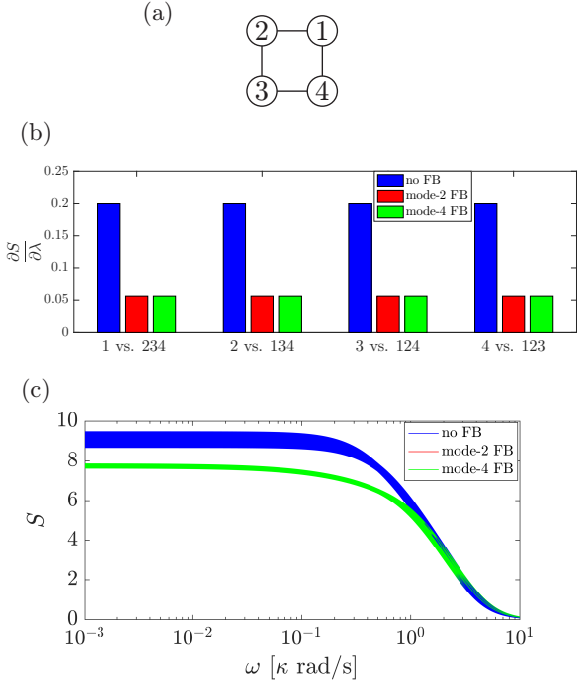}
\caption{
(a) Structure of the square cluster state. 
(b) Sensitivities of the entanglement entropy. 
(c) Entanglement entropy between the mode-1 and the other modes, with (the red and green 
lines) and without (the blue lines) feedback.}
\label{fig:SquareCluster}
\end{center}
\end{figure}

Finally, we examine the square cluster state shown in Fig. \ref{fig:SquareCluster}(a). 
Again we assume $\lambda=\lambda_l$, $\kappa=\kappa_j$, and $\Delta=\Delta_j$. 
The poles are $s=-\kappa/2 \pm i\Delta$, and $s=-\kappa/2 \pm \sqrt{4\lambda^2-\Delta^2}$, 
leading to $\Delta=2\lambda$. 
The controller is a beamsplitter with $\varrho=0.04$. 
Then with the same parameters choice as in the previous case, the sensitivities and the 
entanglement entropies are depicted in Figs. \ref{fig:SquareCluster}(b) and (c). 
As expected, the mode-1 FB and the mode-2 FB schemes have the same effect on the 
robustness, due to the symmetric structure.

%%%%%%%%%%%%%%%%%%%%%%%%%%%%%%%%%%%%%%%%%%%%%%%%
%%%%%%%%%%%%%%%%%%%%%%%%%%%%%%%%%%%%%%%%%%%%%%%%
%%%%%%%%%%%%%%%%%%%%%%%%%%%%%%%%%%%%%%%%%%%%%%%%

\section{Conclusion}

This paper has demonstrated that the feedback amplification technique proposed in 
\cite{Yamamoto2016} is effective for generating robust Gaussian entangled states. 
In particular, we have seen that the degree of robustness depends on the structure of 
feedback control; 
for the four-mode cluster states examined in this paper, our conclusion was that we 
should choose the mode having the biggest number of connection to the others, to 
construct the feedback loop. 
However, determining the most effective feedback for the general case is not a trivial 
problem and needs extensive investigation. 
Considering the fact that a feedback amplification architecture is involved in 
almost all electric circuits to generate robust functionalities, therefore, the result shown 
in this paper would be a first step toward developing a quantum circuit theory for robust 
quantum functionalities such as teleportation.

%%%%%%%%%%%%%%%%%%%%%%%%%%%%%%%%%%%%%%%%%%%%%%%%
%%%%%%%%%%%%%%%%%%%%%%%%%%%%%%%%%%%%%%%%%%%%%%%%
%%%%%%%%%%%%%%%%%%%%%%%%%%%%%%%%%%%%%%%%%%%%%%%%

\appendix

\section{Entanglement Measure}
\label{sec:EntanglementMeasure}

Let us consider a density operator ${\rho}$ of a total system $\mathcal{H}_{\mathrm{tot}}$, 
which we can divide it into two Hilbert spaces: 
$\mathcal{H}_{\mathrm{tot}}=\mathcal{H}_A \otimes \mathcal{H}_B$. 
Then, the entanglement entropy $S_A$ of the Hilbert space $\mathcal{H}_A$ is defined as 
$S_A:=-\Tr_A [{\rho}_A \ln {\rho}_A]$, where ${\rho}_A$ is the reduced density operator 
${\rho}_A=\Tr_B[{\rho}]$. 
It is worth noting that the entropy always satisfies $S_A\geq 0$ where $S_A=0$ means 
that $\mathcal{H}_A$ and $\mathcal{H}_B$ are not entangled. 
Also, it satisfies $S_A=S_B$ iff the state in the total system $\mathcal{H}_{\mathrm{tot}}$ 
is in a pure state.

Although the entanglement entropy is generally hard to calculate, it is straightforward in the 
Gaussian case \cite{Demarie2012, Weedbrook2011, Adesso2014, Serafini2004}. 
Let us consider an $n$-mode Gaussian system with vector of quadratures 
${\mathrm{ \bm{x} }}=[{q}_1,p_1, q_2,p_2,\cdots, {q}_n, {p}_n]^\top$, with 
${q}=({b}+{b}^\dag)/\sqrt2$ and ${p}=({b}-{b}^\dag)/\sqrt2 i$. 
Then the covariance matrix (CM) $\gamma$ is defined as 
\begin{equation*}
        \gamma
           :=\Re\left[ \left\langle {\mathrm{ \bm{x} }} {\mathrm{ \bm{x} }}^\top \right\rangle \right], 
\end{equation*}
where $\langle \bullet \rangle$ is an expected value. 
Let us denote ${\mathrm{ \bm{x} }}_{ \mathrm{out} }$ and ${\mathrm{ \bm{x} }}_{ \mathrm{in} }$ 
as vectors of the output and the input modes respectively. 
These two vectors are related by 
${\mathrm{ \bm{x} }}_{ \mathrm{out} }=Y {\mathrm{ \bm{x} }}_{ \mathrm{in} }$, where 
$Y$ is a $2n\times 2n $ matrix. 
In the vacuum input case, where $\langle q^2 \rangle=\langle p^2 \rangle=1/2$, 
$\langle qp \rangle=i/2$, and $\langle pq \rangle=-i/2$ hold, the CM of the output state is given by 
\[
        \gamma 
        = \Re\left[ Y \langle \mathrm{\bm{x}}_{\mathrm{in}}\mathrm{\bm{x}}_{\mathrm{in}}^\top \rangle 
              Y^\top \right] 
        = \dfrac12 \Re[ YRY^\top ], 
\]
where 
\[
R:=\left[
\begin{array}{cc}
1 &i  \\
-i &1 
\end{array}
\right]\oplus 
\left[
\begin{array}{cc}
1 &i  \\
-i &1 
\end{array}
\right].
\]
Now, the entanglement entropy between the $j$th mode and all the other modes is 
given by 
\[
     S_j = \Big( \sigma_j+\dfrac{1}{2} \Big) \ln \Big( \sigma_j+\dfrac{1}{2} \Big)
                 -\Big( \sigma_j-\dfrac{1}{2} \Big)\ln \Big( \sigma_j-\dfrac{1}{2} \Big),
\]
where $\sigma_j$ is the $j$th symplectic eigenvalue of the CM. 
More specifically, $\pm i\sigma_j$ is the eigenvalue of $\gamma_j \Omega$, where 
\begin{equation*}
      \gamma_j
          =\left[\begin{array}{cc}
              \langle q_j^2 \rangle & \langle q_jp_j + p_jq_j \rangle/2 \\
              \langle q_jp_j + p_jq_j \rangle/2 & \langle p_j^2 \rangle
            \end{array}\right],~
     \Omega=\left[\begin{array}{cc}
                       0 & 1 \\
                       -1 &0
                    \end{array}\right]. 
\end{equation*}
Note that if $\sigma_j=1/2$ then $S_j=0$, meaning that there is no entanglement 
between the mode-$j$ and the others.

%%%%%%%%%%%%%%%%%%%%%%%%%%%%%%%%%%%%%%%

\section{NDPO dynamics for the four mode Gaussian cluster state}
\label{sec:ClusterAppendix}

%\subsection{Linear cluster state}

\begin{figure}[t]
\centering
\includegraphics[width=6cm]{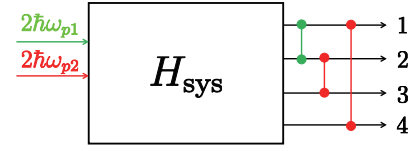}%
\caption{Schematic diagram of the NDPO generating the linear Gaussian cluster state. 
The red and the green lines represent entanglement between the modes.}
\label{fig:GeneratingLinear}
\end{figure}

The linear cluster state is generated, by using two pumps entering a nonlinear crystal 
(see Fig.~\ref{fig:GeneratingLinear}). 
The pump with frequency $2\omega_{p1}$ generates a pair of the modes $\{ 1, 2\}$, and 
the pump with frequency $2\omega_{p2}$ generates two pairs of the modes $\{ 2,3 \}$ 
and $\{ 1,4 \}$. 
That is, the system Hamiltonian is given by
\begin{align*}
    H_{\mathrm{sys}}=\sum_{j=1}^4 \omega_j a_j^\dag a_j 
           &+ \Big[i\big( \eta_1 a_1^\dag a_2^\dag a_{p1} + \eta_2 a_1^\dag a_4^\dag a_{p2} \\
                         & + \eta_3a_2^\dag a_3^\dag a_{p2} \big) + {\rm h.c.} \Big],
\end{align*}
where $a_j$ is the annihilation operator of mode-$j$. 
Also $\omega_j$ is the frequency of mode-$j$, and $\eta_l$ is the coupling constant between 
the cavity modes. 
\begin{figure*}[t]
\begin{center}
    \includegraphics[clip,width=13cm]{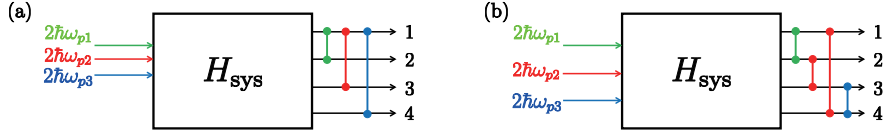}
\caption{
Schematic diagrams of the NDPO generating (a) the T-shape and (b) the square cluster states.}
\label{fig:GeneratingTshapeAndSquare}
\end{center}
\end{figure*}
We now use the undepleted pump approximation 
\cite{Analysis053826, Quadripartite063834, Multipartite020302} to make the interactions linear; 
that is, $a_{p1}$ and $a_{p2}$ are treated as c-numbers and are absorbed into $\eta_l$ which 
redefines the coupling constants as $\eta_l\rightarrow \lambda_l$. 
The resulting quantum Langevin equations for the cavity modes $a_j$ are
\begin{align}
\dot{a}_1&=-\left( i\Delta_1 + \dfrac{\kappa_1}{2} \right)a_1+ \lambda_1 a_2^\dag + \lambda_2 a_4^\dag -\sqrt{\kappa_1}b_{1,\mathrm{in}}, \notag \\
\dot{a}_2^\dag&=-\left( -i\Delta_2 + \dfrac{\kappa_2}{2} \right)a_2^\dag + \lambda_1 a_1 + \lambda_3 a_3 -\sqrt{\kappa_2}b_{2,\mathrm{in}}^\dag,\notag \\
\dot{a}_3&=-\left( i\Delta_3 + \dfrac{\kappa_3}{2} \right)a_3 + \lambda_3 a_2^\dag -\sqrt{\kappa_3}b_{3,\mathrm{in}},\notag \\
\dot{a}_4^\dag&=-\left( -i\Delta_4 + \dfrac{\kappa_4}{2} \right)a_4^\dag + \lambda_2 a_1 -\sqrt{\kappa_4}b_{4,\mathrm{in}}^\dag,\notag
\end{align} 
where $\kappa_j$ is the damping rates, $\Delta_j$ is the detuning, and $b_{j,\mathrm{in}}$ 
is the input field mode interacting with $a_j$. 
Also, the input-output relations are given by 
$b_{j,\mathrm{out}}=b_{j,\mathrm{in}}+\sqrt{\kappa_j}a_j$. 
After the Laplace transform is applied to the above quantum Langevin equations, the input 
fields $b_{j,\mathrm{in}}$ and the output fields $b_{j,\mathrm{out}}$ are related by the 
following symmetric transfer function matrix $G(s)=G(s)^\top$: 
\begin{align}
\left[
\begin{array}{c}
b_1  \\
b_{2}^\dag  \\
b_{3}  \\
b_{4}^\dag 
\end{array}
\right]_{\mathrm{out}}
=\left[
\begin{array}{cccc}
G_{11} &G_{12} &G_{13} &G_{14}  \\
\star &G_{22} &G_{23} &G_{24}  \\
\star & \star &G_{33} &G_{34}  \\
\star & \star & \star &G_{44} 
\end{array}
\right]
\left[
\begin{array}{c}
b_{1}  \\
b_{2}^\dag  \\
b_{3}  \\
b_4^\dag 
\end{array}
\right]_{\mathrm{in}}, \notag
\end{align} 
where $\star$ denote the symmetric elements and 
\begin{align}
G_{11}&=1+ \kappa_1 A_4 (A_2 A_3 -\lambda_3^2)/D,~
G_{12}=\lambda_1 \sqrt{\kappa_1 \kappa_2}A_3 A_4/D, \notag \\
G_{13}&=\lambda_1 \lambda_3 \sqrt{\kappa_1 \kappa_3} A_4/D,~
G_{14}=\lambda_2 \sqrt{\kappa_1 \kappa_4} (A_2 A_3 - \lambda_3^2)/D,\notag \\
G_{22}&=1+  \kappa_2 A_3 (A_1 A_4 -\lambda_2^2) /D,\notag \\
G_{23}&=\lambda_3 \sqrt{\kappa_2 \kappa_3} (A_1 A_4 -\lambda_2^2)/D,~
G_{24}=\lambda_1 \lambda_2 \sqrt{\kappa_2 \kappa_4} A_3/D,\notag \\
G_{33}&=1+ \kappa_3 (A_1 A_2 A_4 -\lambda_1^2 A_4 -\lambda_2^2 A_2) /D,\notag \\
G_{34}&=\lambda_1 \lambda_2 \lambda_3 \sqrt{\kappa_3 \kappa_4}/D, \notag \\
G_{44}&=1+  \kappa_4 (A_1 A_2 A_3 -\lambda_1^2 A_3 -\lambda_3^2 A_1) /D,\notag
\end{align} 
where 
\begin{align}
     D&=\lambda_1^2 A_3A_4-(A_1A_4 - \lambda_2^2)(A_2 A_3 - \lambda_3^2), \notag \\
     A_1&=s+i\Delta_1+\kappa_1/2,~A_2=s-i\Delta_2 + \kappa_2/2,\notag \\
     A_3&=s+i\Delta_3 + \kappa_3/2,~A_4=s-i\Delta_4+\kappa_4/2.\notag
\end{align}
The $Y$ matrix appearing in Appendix~\ref{sec:EntanglementMeasure} is readily obtained from $G$, and as a result 
the entanglement entropy of the linear cluster state can also be calculated easily.

As for the T-shape and square cluster states, we just give the system Hamiltonians. 
In these cases three pump modes are used, as seen in Fig.~\ref{fig:GeneratingTshapeAndSquare}. 
For the T-shape case, the system Hamiltonian within the undepleted pump approximation is 
given by 
\begin{align*}
    H_{\mathrm{sys}}
         =\sum_{j=1}^4 \omega_j a_j^\dag a_j 
                 + \Big[ i \big( \lambda_1a_1^\dag a_2^\dag +\lambda_2 a_1^\dag a_3^\dag 
                 + \lambda_3 a_1^\dag a_4^\dag \big) + {\rm h.c.} \Big] \notag
\end{align*} 
Also for the square cluster state, it is given by 
\begin{align*}
   H_{\mathrm{sys}}
       =\sum_{j=1}^4 \omega_j a_j^\dag a_j 
          &+ \Big[ i \big( \lambda_1a_1^\dag a_2^\dag + \lambda_2 a_2^\dag a_3^\dag \\
          &+ \lambda_3 a_1^\dag a_4^\dag + \lambda_4 a_3^\dag a_4^\dag \big) + {\rm h.c.} \Big]
\end{align*} 
%

%%%%%%%%%%%%%%%%%%%%%%%%%%%%%%%%%%%%%%%%%%%%%%%%
%%%%%%%%%%%%%%%%%%%%%%%%%%%%%%%%%%%%%%%%%%%%%%%%
%%%%%%%%%%%%%%%%%%%%%%%%%%%%%%%%%%%%%%%%%%%%%%%%

%\end{comment}

%\bibliography{NDPObiblio}

\end{document}